\documentclass{llncs}
\usepackage[fleqn]{amsmath}
\usepackage[T1]{fontenc} 
\usepackage{graphicx}
\usepackage{xcolor}
\usepackage{algorithm}
\usepackage{algorithmic}
\usepackage{color,soul,url}
\usepackage{adjustbox}
\usepackage{booktabs}
\usepackage{lipsum}
\usepackage{multirow}
\usepackage{makecell}
\usepackage{caption}
\usepackage{subcaption}
\usepackage{tabularx}

\begin{document}

\mainmatter
\title{Less is More: Robust and Novel Features for Malicious Domain Detection}
\titlerunning{Robust Malicious Domain Detection} 

%

\author{Chen Hajaj\inst{1} \and
Nitay Hason\inst{2} \and
Nissim Harel \inst{3}\and
Amit Dvir\inst{4}
}
\authorrunning{Chen Hajaj et al.}
%

%
\institute{
	Ariel Cyber Innovation Center, Data Science and Artificial Intelligence Research Center, IEM Department, Ariel University,  Israel\\
	\email{chenha@ariel.ac.il}\\
	\texttt{https://www.ariel.ac.il/wp/chen-hajaj/}
	\and
	Ariel Cyber Innovation Center, CS Department, Ariel University, Israel\\
	\email{nitay.has@gmail.com}
	\and
	CS Department, Holon Institute of Technology, Israel\\
	\email{nissimh@hit.ac.il}
	\and
	Ariel Cyber Innovation Center, CS Department, Ariel University, Israel\\
	\email{amitdv@ariel.ac.il}\\
	\texttt{https://www.ariel.ac.il/wp/amitd/}}
\maketitle

\begin{abstract}
Malicious domains are increasingly common and pose a severe cybersecurity threat. Specifically, many types of current cyber attacks use URLs for attack communications (e.g., C\&C, phishing, and spear-phishing). Despite the continuous progress in detecting these attacks, many alarming problems remain open, such as the weak spots of the defense mechanisms. Since machine learning has become one of the most prominent methods of malware detection, A robust feature selection mechanism is proposed that results in malicious domain detection models that are resistant to evasion attacks. This mechanism exhibits high performance based on empirical data. This paper makes two main contributions: First, it provides an analysis of robust feature selection based on widely used features in the literature. Note that even though the feature set dimensional space is reduced by half (from nine to four features), the performance of the classifier is still improved (an increase in the model's F1-score from 92.92\% to 95.81\%). Second, it introduces novel features that are robust to the adversary's manipulation. Based on extensive evaluation of the different feature sets and commonly used classification models this paper show that models which are based on robust features are resistant to malicious perturbations, and at the same time useful for classifying non-manipulated data.

\end{abstract}

\section{Introduction} 
\label{Chapter1} 


\newcommand{\keyword}[1]{\textbf{#1}}
\newcommand{\tabhead}[1]{\textbf{#1}}
\newcommand{\code}[1]{\texttt{#1}}
\newcommand{\file}[1]{\texttt{\bfseries#1}}
\newcommand{\option}[1]{\texttt{\itshape#1}}

\begin{figure}[htp]
	\centering
	\includegraphics[width=8cm]{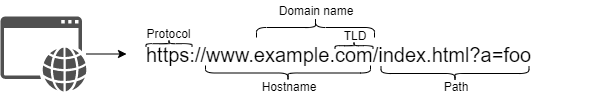}
	\caption{The URL structure}
	\label{fig:webdomainparts}
\end{figure}

\setcounter{footnote}{0} 
In the past two decades, cybersecurity attacks have become a major issue for governments and civilians~\cite{shu2017breaking}. Many of these attacks are based on malicious web domains or URLs (See Figure~\ref{fig:webdomainparts} for the structure of a URL). These domains are used for phishing~\cite{blum2010lexical,khonji2013phishing,le2011phishdef,prakash2010phishnet,sheng2009empirical} (e.g. spear phishing), Command and Control (C\&C)~\cite{sandell1978survey} and a vast set of virus and malware~\cite{canali2011prophiler} attacks.

The Domain Name System (DNS) maps human-readable domain names to their associated IP addresses (e.g., google.com to 172.217.16.174). However, DNS services are abused in different ways in order to conduct various attacks. In such attacks, the adversary can utilize a set of domains and IP addresses to orchestrate sophisticated attacks~\cite{shi2017malicious,antonakakis2010building}. 
Therefore, the ability to identify a malicious domain in advance is a massive game-changer~\cite{ahmed2018fault,antonakakis2010building,wrinkle,bilge2014exposure,caglayan2009real,choi2011detecting,dolberg2012efficient,MiSAL,hu2016identifying,huang2006extreme,nelms2013execscent,peng2018detecting,rahbarinia2016efficient,shi2017malicious,sun2019hindom,torabi2018detecting,yadav2012detecting}. 

A common way of identifying malicious/compromised domains is to collect information about the domain names (alphanumeric characters) and network information (such as DNS and passive DNS data\footnote{Most works dealing with malicious domain detection are based on DNS features, and only some take the passive DNS features into account as well.}). This information is then used for extracting a set of features, according to which machine learning (ML) algorithms, are trained based on a desirably massive amount of data~\cite{antonakakis2010building,antonakakis2011detecting,wrinkle,bilge2014exposure,caglayan2009real,choi2011detecting,MiSAL,hu2016identifying,huang2006extreme,nelms2013execscent,peng2018detecting,perdisci2012early,rahbarinia2016efficient,sun2019hindom,yadav2012detecting}. 

A mathematical approach can also be used in a variety of ways~\cite{dolberg2012efficient,yadav2012detecting}, such as measuring the distance between a known malicious domain name and the analyzed domain (benign or malicious)~\cite{yadav2012detecting}. Still, while ML-based solutions are widely used, many of them are not robust; an attacker can easily bypass these models with minimal feature perturbations (e.g., change the length of the domain or modify network parameters such as Time To Live, TTL)~\cite{Papernot16,tong2019framework}. 

In this context, one of the main questions is how to identify malicious/compromised domains in the presence of an intelligent adversary that can manipulate domain properties.

For these reasons, a feature selection mechanism which is robust to adversarial manipulations is used to tackle the problem of identifying malicious domains. Thus, even if the attacker has a black-box access to the model, tampering with the domain properties or network parameters will have a negligible effect on the model's accuracy. In order to achieve this goal, a broad set of both malicious and benign URLs were collected and surveyed for commonly used features. These features were then manipulated to show that some of them, although widely used, are only slightly robust or not robust at all. Thus, novel features were engineered to enhance the robustness of the models. While these novel features support the detection of malicious domains from new angles (e.g. autonomous system numbers), the resulting accuracy of models that were solely based on these features is not necessarily higher than the original features. Therefore, a hybrid set of features was generated, combining a subset of the well-known features, with the novel features.  Finally, the different sets of features were evaluated using well-known machine learning and deep learning algorithms, which led to a robust and efficient malicious domain detection system.

The rest of the paper is organized as follows: Section \ref{sec:evl} presents the evaluation metric used, and Section \ref{Chapter3} summarizes related works. Section \ref{Chapter4} describes the methodology and the novel features. Section \ref{Chapter5} presents the empirical analysis and evaluation. Finally, Section \ref{Chapter6} concludes and summarizes this work.

\section{Evaluation Metrics}
\label{sec:evl}
Machine Learning (ML) is a subfield of Computer Science aimed at getting computers to act and improve over time autonomously by feeding them data in the form of observations and real-world interactions. In contrast to traditional programming, when one provides input and algorithm and receives an output when using ML, one provides a list of inputs and their associated outputs, in order to extract the algorithm that maps the two.
 
ML algorithms are often categorized as either supervised or unsupervised. In supervised learning, each example is a pair consisting of an input vector (also called data point) and the desired output value (class/label). Unsupervised learning learns from data that has not been labeled, classified, or categorized. Instead of responding to feedback, unsupervised learning identifies commonalities in the data and reacts based on the presence or absence of such commonalities in each new piece of data. 

In order to evaluate how a supervised model is adapted to a problem, the dataset needs to be split into two, the training set and the testing set. The training set is used to train the model, and the testing set is used to evaluate how well the model "learned" (i.e. by comparing the model predictions with the known labels). Usually, the train/test distribution is around 75\%/25\% (depending on the problem and the amount of data). Standard evaluation criteria are as follows: Recall, Precision, Accuracy, F1-score, and Loss. All of these criteria can easily be extracted from the evaluation's confusion matrix.

Confusion matrix (Table~\ref{tab:confusion_matrix}) is commonly used to describe the performance of a classification model. WLOG, we define positive instances as malicious and negative as benign. Recall (Eq.~\ref{ml:recall}) is defined as the number of correctly classified malicious examples out of all the malicious ones. Similarly, Precision (Eq.~\ref{ml:precision}) is the number of correctly classified malicious examples out of all examples classified as malicious (both correctly and wrongly classified). Accuracy (Eq.~\ref{ml:accuracy}) is used as a statistical measure of how well a classification test correctly identifies or excludes a condition. That is, the accuracy is the proportion of true results (both true positives and true negatives) among the total number of cases examined. Finally, F1-score (Eq.~\ref{ml:f1_score}) is a measure of a test's accuracy. It considers both the precision and the recall of the test to compute the score. The F1-score is the harmonic average of the precision and recall, where an F1-score reaches its best value at 1 (perfect precision and recall) and worst at 0. These criteria are used as the main evaluation metric. Since a classification model is being examined here, the Logarithmic loss (Eq.~\ref{ml:log_loss}) was chosen as the loss function. 

In this research, the problem of identifying malicious web domains can be classified as supervised learning, as the correct label (i.e. malicious or benign) can be extracted using a blacklist-based method, as we describe in the next chapter.

\begin{table}[]
\begin{tabularx}{\textwidth}{llcclllllllll}
\multicolumn{1}{c}{}                                                                                   &                                     & \multicolumn{1}{l}{}                                                          & \multicolumn{1}{l}{}                                                          &                                     &  &  &  &  &  &  &  &  \\ \cline{3-5}
                                                                                                       & \multicolumn{1}{l|}{}               & \multicolumn{2}{l|}{\textbf{Prediction Outcome}}                                                                                                             & \multicolumn{1}{l|}{}               &  &  &  &  &  &  &  &  \\ \cline{3-5}
                                                                                                       & \multicolumn{1}{l|}{}               & \multicolumn{1}{c|}{Positive}                                                 & \multicolumn{1}{c|}{Negative}                                                 & \multicolumn{1}{l|}{\textbf{Total}} &  &  &  &  &  &  &  &  \\ \cline{1-5}
\multicolumn{1}{|c|}{\multirow{2}{*}{\textbf{\begin{tabular}[c]{@{}c@{}}Actual\\ Value\end{tabular}}}} & \multicolumn{1}{l|}{Positive}       & \multicolumn{1}{c|}{\begin{tabular}[c]{@{}c@{}}True\\ Positive\end{tabular}}  & \multicolumn{1}{c|}{\begin{tabular}[c]{@{}c@{}}False\\ Negative\end{tabular}} & \multicolumn{1}{l|}{TP+FN}          &  &  &  &  &  &  &  &  \\ \cline{2-5}
\multicolumn{1}{|c|}{}                                                                                 & \multicolumn{1}{l|}{Negative}       & \multicolumn{1}{c|}{\begin{tabular}[c]{@{}c@{}}False\\ Positive\end{tabular}} & \multicolumn{1}{c|}{\begin{tabular}[c]{@{}c@{}}True\\ Negative\end{tabular}} & \multicolumn{1}{l|}{FP+TN}          &  &  &  &  &  &  &  &  \\ \cline{1-5}
\multicolumn{1}{|l|}{}                                                                                 & \multicolumn{1}{l|}{\textbf{Total}} & \multicolumn{1}{c|}{P}                                                        & \multicolumn{1}{c|}{N}                                                        & \multicolumn{1}{l|}{}               &  &  &  &  &  &  &  &  \\ \cline{1-5}
\end{tabularx}
    \caption{Confusion Matrix}
    \label{tab:confusion_matrix}
\end{table}

    \begin{equation}\label{ml:recall}
    Recall = \frac{TP}{TP + FN}
    \end{equation}
    \begin{equation}\label{ml:precision}
    Precision = \frac{TP}{TP + FP}=\frac{TP}{P}
    \end{equation}
    \begin{equation}\label{ml:accuracy}
    Accuracy = \frac{TP + TN}{TP + FP + TN + FN}=\frac{T}{P+N}
    \end{equation}
    \begin{equation}\label{ml:f1_score}
    F_{1}-score = 2\cdot\frac{Precision\cdot Recall}{Precision + Recall}
    \end{equation}
    \begin{equation}\label{ml:log_loss}
    Loss = -\frac{1}{N}\sum^{N}_{i=1}(y_{i}\cdot \log(p_{i})+(1-y_{i})\cdot \log(1-p_{i}))
    \end{equation}

\section{Related Work} 
\label{Chapter3}
The issue of identifying malicious domains is a fundamental problem in cybersecurity. This section discusses recent results in identifying malicious domains, focusing on three significant methodologies: Mathematical Theory approaches, ML-based techniques, and Big Data approaches.

The use of graph theory to identify malicious domains was more pervasive in the past~\cite{dolberg2012efficient,jung2004empirical,mishsky2015topology,othman2017advanced,yadav2012detecting}.  Yadav et al.~\cite{yadav2012detecting} presented a method for recognizing malicious domain names based on fast flux. Fast flux is a DNS technique used by botnets to hide phishing and malware delivery sites behind an ever-changing network of compromised hosts acting as proxies. Their methodology analyzed the DNS queries and responses to detect if and when domain names are being generated by a Domain Generation Algorithm (DGA). Their solution was based on computing the distribution of alphanumeric characters for groups of domains and by statistical metrics with the KL (Kullback Leibler) distance, Edit distance and Jaccard measure to identify these domains.
Their results for a fast-flux attack using the Jaccard Index achieved impressive results, with 100\% detection and 0\% false positives. However, for smaller numbers of generated domains for each TLD, their false positive results were much higher, at 15\% when 50 domains were generated for the TLD using the KL-divergence over unigrams, and 8\% when 200 domains were generated for each TLD using Edit distance.
    


Dolberg et al.~\cite{dolberg2012efficient} described a system called \emph{Multi-dimensional Aggregation Monitoring (MAM)} that detects anomalies in DNS data by measuring and comparing a “steadiness” metric over time for domain names and IP addresses using a tree-based mechanism. The steadiness metric is based on a similar domain to IP resolution patterns when comparing DNS data over a sequence of consecutive time frames. The domain name to IP mappings were based on an aggregation scheme and measured steadiness. In terms of detecting malicious domains, the results showed that an average steadiness value of 0.45 could be used as a reasonable threshold value, with a 73\% true positive rate and only 0.3\% false positives. The steadiness values might not be considered a good indicator when fewer malicious activities were present (e.g. $ < $10\%).

However, the most common approach to identifying malicious domains is using machine learning (ML)~\cite{antonakakis2010building,antonakakis2011detecting,caglayan2009real,dasmachine2019,nelms2013execscent,perdisci2012early,sahoo2017malicious,shi2017malicious,sun2019hindom}. Using a set of extracted features, researchers can train ML algorithms to label URLs as malicious or benign. Shi et al.~\cite{shi2017malicious} proposed a machine learning methodology to detect malicious domain names using the Extreme Learning Machine (ELM)~\cite{huang2006extreme} which is closest to the one employed here. ELM is a new neural network with high accuracy and fast learning speed. The authors divided their features into four categories: construction-based, IP-based, TTL-based, and WHOIS-based. Their evaluation resulted in a high detection rate, an accuracy exceeding 95\%, and a fast learning speed. However, as shown below, a significant fraction of the features used in this work emerged as non-robust and ineffective in the presence of an intelligent adversary.

Sun et al.~\cite{sun2019hindom} presented a system called \emph{HinDom}, that generate a heterogeneous graph (in contrast to homogeneous graphs created by~\cite{yadav2012detecting,rahbarinia2016efficient}) in order to robustly identify malicious attacks (e.g. spams, phishing, malware and botnets). Even though HinDom collected DNS and pDNS data, it also has the ability to collect information from various clients inside networks (e.g. CERNET2 and TUNET) and by that the perspective of it is different from the perspective of this study (i.e. client perspective). Nevertheless, HinDom has achieved remarkable results using transductive classifier it managed to achieve high accuracy and F1-score with 99\% and 97.5\% respectively.

Bilge et al.~\cite{bilge2014exposure} created a system called \emph{Exposure},  designed to detect malicious domain names. Their system uses passive DNS data collected over some period of time to extract features related to known malicious and benign domains. Passive DNS Replication\cite{torabi2018detecting,antonakakis2010building,antonakakis2011detecting,bilge2014exposure,perdisci2012early,nelms2013execscent,rahbarinia2016efficient} refers to the reconstruction of DNS zone data by recording and aggregating live DNS queries and responses. Passive DNS data could be collected without requiring the cooperation of zone administrators. The Exposure is designed to detect malware- and spam-related domains. It can also detect malicious fast-flux and DGA-related domains based on their unique features. The system computes the following four sets of features from anonymized DNS records: (a) Time-based features related to the periods and frequencies that a specific domain name was queried in; (b) DNS-answer-based features calculated based on the number of distinctive resolved IP addresses and domain names, the countries that the IP addresses reside in, and the ratio of the resolved IP addresses that can be matched with valid domain names and other services; (c) TTL-based features that are calculated based on statistical analysis of the TTL over a given time series; (d) Domain name-based features are extracted by computing the ratio of the numerical characters to the domain name string, and the ratio of the size of the longest meaningful substring in the domain name. Using a Decision Tree model, Exposure reported a total of 100,261 distinct domains as being malicious, which resolved to 19,742 unique IP addresses. The combination of features that were used to identify malicious domains led to the successful identification of several domains that were related to botnets, flux networks, and DGAs, with low false positive and high detection rates. It may not be possible to generalize the detection rate results reported by the authors (98\%) since they were highly dependent on comparisons with biased datasets. Despite the positive results, once an identification scheme is published, it is always possible for an attacker to evade detection by mimicking the behaviors of benign domains.

Rahbarinia et al.~\cite{rahbarinia2016efficient} presented a system called \emph{Segugio}, an anomaly detection system based on passive DNS traffic to identify malware-controlled domain names based on their relationship to known malicious domains. The system detects malware-controlled domains by creating a machine domain bipartite graph that represents the underlying relations between new domains and known benign/malicious domains. The system operates by calculating the following features: (a) Machine Behavior, based on the ratio of “known malicious” and “unknown” domains that query a given domain d over the total number of machines that query d. The larger the total number of queries and the fraction of malicious related queries, the higher the probability that d is a malware controlled domain; (b) Domain Activity, where given a time period, domain activity is computed by counting the total number of days in which a domain was actively queried; (c) IP Abuse, where given a set of IP addresses that the domain resolves to, this feature represents the fraction of those IP addresses that were previously targeted by known malware controlled domains. Using a Random Forest model, Segugio was shown to produce high true positive and very low false positive rates (94\% and 0.1\% respectively). It was also able to detect malicious domains earlier than commercial blacklisting websites. However, Segugio is a system that can only detect malware related domains based on their relationship to previously known domains and therefore cannot detect new (unrelated to previous malicious domains) malicious domains. More information about malicious domain filtering and malicious URL detection can be found in~\cite{dasmachine2019,sahoo2017malicious}.


Adversarial Machine Learning is a subfield of Machine Learning in which the training and testing set do not share the same distribution, for example, given perturbations on a malicious instance so that it will be falsely classified. These manipulated instances are commonly called \textit{adversarial examples (AE)}\cite{goodfellow2014explaining}. AE are samples an attacker changes, based on some knowledge of the model classification function. These examples are slightly different from correctly classified examples. Therefore, the model fails to classify them correctly. AE are widely used in the fields of spam filtering\cite{nelson2008exploiting}, network intrusion detection systems (IDS)\cite{fogla2006polymorphic}, Anti-Virus signature tests\cite{newsome2006paragraph} and  bio-metric recognition\cite{rodrigues2009robustness}. 

Attackers commonly follow one of two models to generate adversarial examples: 1) white-box attacker\cite{madry2017towards,raghunathan2018certified,song2017pixeldefend}, which has full knowledge of the classifier and the train/test data; 2) black-box attacker\cite{madry2017towards,papernot2017practical,shahpasand2019adversarial}, which has access to the model's output for each given input. 

Various methods emerged to tackle AE-based attacks and make ML models robust. The most promising are those based on game-theoretic approaches\cite{bruckner2011stackelberg,singh2011game,zolotukhin2013support}, robust optimization\cite{madry2017towards,raghunathan2018certified,xu2009robustness}, and adversarial retraining\cite{li2018evasion,nissim2016aldroid,tong2019framework}. These approaches mainly concern \textit{feature-space models} of attacks where feature space models assume that the attacker changes values of features directly. Note that these attacks may be an abstraction of reality as random modifications to feature values may not be realizable or avoid the manipulated instance functionality.

Big Data is an evolving term that describes any voluminous amount of structured, semi-structured and unstructured data that can be mined for information. Big data is often characterized by 3Vs: the extreme Volume of data, the wide Variety of data types and the Velocity at which the data must be processed. To implement Big Data, high volumes of low-density, unstructured data need to be processed. This can be data of unknown value, such as Twitter data feeds, click streams on a web page or a mobile app, or sensor enabled equipment. For some organizations, this might be tens of terabytes of data. For others, it may be hundreds of petabytes. Velocity is the fast rate at which data are received and (perhaps) acted on. Normally, the highest velocity of data streams directly into memory rather than being written to disk. 

Torabi et al.~\cite{torabi2018detecting} surveyed state of the art systems that utilize passive DNS traffic for the purpose of detecting malicious behaviors on the Internet. They highlighted the main strengths and weaknesses of these systems in an in depth analysis of the detection approach, collected data, and detection outcomes. They showed that almost all systems have implemented supervised machine learning. In addition, while all these systems require several hours or even days before detecting threats, they can achieve enhanced performance by implementing a system prototype that utilizes big data analytic frameworks to detect threats in near real-time. This overview contributed in four ways to the literature. (1) They surveyed implemented systems that used passive DNS analysis to detect DNS abuse/misuse; (2) they performed an in-depth analysis of the systems and highlighted their strengths and limitations; (3) they implemented a system prototype for near real-time threat detection using a big data analytic framework and passive DNS traffic; (4) they presented real-life cases of DNS misuse/abuse to demonstrate the feasibility of a near real time threat detection system prototype. However, the cases that were presented were too specific. In order to understand the real abilities of their system, the system must be analyzed with a much larger test dataset.

\section{Methodology}
\label{Chapter4}
The following criteria are used: Section~\ref{section:data_collecting} outlines the characteristics and methods of collection of the dataset. Section ~\ref{section:feature_engineering}, defines each of the well known features evaluated. Section~\ref{sec:robust} covers the evaluation of their robustness, and Section~\ref{sec:new_fea} presents novel features and evaluates their robustness.

\subsection{Data Collection}
\label{section:data_collecting}
The main ingredient of ML models is the data on which the models are trained. As discussed above, data collection should be as heterogeneous as possible to model reality. The data collected for this work include both malicious and benign URLs: the benign URLs are based on the Alexa top 1 million~\cite{alexasite}, and the malicious domains were crawled from multiple sources~\cite{phishtanksite,urlhaussite,scumwaresite} due to the fact they are quite rare. 

According to~\cite{webrootsite}, 25\% of all URLs in 2017 were malicious, suspicious or moderately risky. Therefore, to make a realistic dataset, all the evaluations include all 1,356 malicious active unique URLs, and consequently 5,345 benign active unique URLs as well.
For each instance, the URL and domain information properties were crawled from Whois, and DNS records. Whois is a widely used Internet record listing that identifies who owns a domain, how to get in contact with them, the creation date, update dates, and expiration date of the domain. Whois records have proven to be extremely useful and have developed into an essential resource for maintaining the integrity of the domain name registration and website ownership. Note that according to a study by ICANN~\footnote{Internet Corporation for Assigned Names and Numbers}~\cite{icannstudy2013}, many malicious attackers abuse the Whois system. Hence, only the information that could not be manipulated was used.

Finally, based on these resources (Whois and DNS records), the following features were generated: the length of the domain, the number of consecutive characters, and the entropy of the domain from the URLs' datasets. Next, the lifetime of the domain and the active time of domain were calculated from the Whois data. Based on the DNS response dataset (a total of 263,223 DNS records), the number of IP addresses, distinct geolocations of the IP addresses, average Time to Live (TTL) value, and the Standard deviation of the TTL were extracted. For extracting the novel features (Section \ref{sec:new_fea}) Virus Total (VT) \cite{virustotalsite} and Urlscan \cite{urlscansite} were used, where Urlscan was used to extract parameters such as the IP address of the page element of the URL.

\subsection{Feature Engineering}
\label{section:feature_engineering}
Based on previous works surveyed, a set of features which are commonly used for malicious domain classification~\cite{antonakakis2010building,antonakakis2011detecting,bilge2014exposure,perdisci2012early,rahbarinia2016efficient,ranganayakulu2013detecting,sahoo2017malicious,shi2017malicious,xiang2011cantina} were extracted. Specifically,the following nine features were used as the baseline:
\begin{itemize}
    \item \textbf{Length of domain}:
    \begin{multline}
           \textrm{Length of domain} = length(Domain_{(i)})
   \end{multline}
    The length of domain is calculated by the domain name followed by the TLD (gTLD or ccTLD). Hence, the minimum length of a domain is four since the domain name needs to be at least one character (most domain names have at least three characters) and the TLD (gTLD or ccTLD) is composed of at least three characters (including the dot character) as well. For example, for the URL http://www.ariel-cyber.co.il, the length of the domain is 17 (the number of characters for the domain name - "ariel-cyber.co.il").
    \\
    \item \textbf{Number of consecutive characters}:
    \begin{multline}
    \textrm{Number of consecutive characters} =\\ \max\{\textrm{consecutive repeated characters in }Domain_{(i)}\}
    \end{multline}
    The maximum number of consecutive repeated characters in the domain. This includes the domain name and the TLD (gTLD or ccTLD). For example for the domain "aabbbcccc.com" the maximum number of consecutive repeated characters value is 4.
    \\
    \item \textbf{Entropy of the domain}:
   \begin{multline}
        \textrm{Entropy of the domain} = \\ -\sum_{j=1}^{n_{i}} \frac{count(c^{i}_{j})}{length(Domain_{(i)})}\cdot \log\frac{count(c^{i}_{j})}{length(Domain_{(i)})}
    \end{multline}
    The calculation of the entropy (i.e. Feature 3) for a given domain $Domain_{(i)}$ consists of $n_{i}$ distinct characters $\{c^{i}_{1},c^{i}_{2},\dots,c^{i}_{n_{i}}\}$. For example, for the domain "google.com" the entropy is\\ $-(5\cdot(\frac{1}{10}\cdot\log\frac{1}{10})+2\cdot(\frac{2}{10}\cdot\log\frac{2}{10})+3(\cdot\frac{3}{10}\cdot\log\frac{3}{10}))=1.25$\\
    The domain has 5 characters that appear once ('l', 'e', '.', 'c', 'm'), one character that appears twice ('g') and one character that appears three times ('o').
    
    \item \textbf{Number of IP addresses}:
    \begin{multline}
           \textrm{Number of IP addresses} = \|\textrm{distinct IP addresses}\|
    \end{multline}
    The number of distinct IP addresses in the domain's DNS record. 
    For example for the list ["1.1.1.1", "1.1.1.1","2.2.2.2"] the number of distinct IP addresses is 2.
    \\
    \item \textbf{Distinct Geo-locations of the IP addresses}:
    \begin{multline}
           \textrm{Distinct Geo-locations of the IP addresses} =\\ \|\textrm{distinct countries}\|
   \end{multline} 
    For each IP address in the DNS record, the countries for each IP were listed and the number of countries was counted. For example for the list of IP addresses ["1.1.1.1", "1.1.1.1","2.2.2.2"] the list of countries is ["Australia", "Australia", "France"] and the number of distinct countries is 2.
    \\
    \item \textbf{Mean TTL value}:
    \begin{multline}
        \textrm{Mean TTL value} = \\ \mu\{\textrm{TTL in DNS records of } Domain_{(i)}\}
    \end{multline}   

    For all the DNS records of the domain in the DNS dataset, the TTL values were averaged. For example, if 30 checks of some domain's DNS records were conducted, and in 20 of these checks the TTL value was "60" and in 10 checks the TTL value was "1200", the mean is $\frac{20\cdot 60+10\cdot 1200}{30}=440$.
    \\
    \item \textbf{Standard deviation of the TTL}:
    \begin{multline}
        \textrm{Standard deviation of TTL} = \\ \sigma\{\textrm{TTL in DNS records of } Domain_{(i)}\}
    \end{multline}
    For all the DNS records of the domain in the DNS dataset, the standard deviation of the TTL values were calculated. For the ''Mean TTL value'' example above, the standard deviation of the TTL values is $537.401$.
    \\
    \item \textbf{Lifetime of domain}:
     \begin{multline}
        \textrm{Lifetime of domain} = \\ Date_{Expiration} - Date_{Created}
    \end{multline}
    \\The interval between a domain's expiration date and creation date in years. For example for the domain "ariel-cyber.co.il", according to Whois information, the dates are: Created on 2015-05-14, Expires in 2020-05-14, Updated on 2015-05-14. Therefore the lifetime of the domain is the number of years from 2015-05-14 to 2020-05-14; i.e. 5.
    \\
    \item \textbf{Active time of domain}:
    \begin{multline}
            \textrm{Active time of domain} = \\ Date_{Updated} - Date_{Created}
    \end{multline}
    Similar to the lifetime of a domain, the active time of a domain is calculated as the interval between a domain's update date and creation date in years. Using the same example as for the ''Lifetime of domain'', the active time of the domain "ariel-cyber.co.il" is the number of years from 2015-05-14 to 2018-06-04; i.e., 3.
\end{itemize}

\subsection{Robust Feature Selection}\label{sec:robust}
Next the robustness of the set of features described above was evaluated to filter those that could significantly harm the classification process given basic manipulations. In the following analysis, the robustness of these features' is analysed (i.e. the complexity of manipulating the feature's values to result in a false classification). Table~\ref{table:base_features} lists the common features along with the mean value and standard deviation for malicious and benign URLs based on the dataset. Strikingly, the table shows that some of the features have similar mean values for both benign and malicious instances. For example, whereas ``Distinct geolocations of the IP addresses'' is quite similar for both types of instances (i.e. not effective in malicious domain classification), it is widely used~\cite{shi2017malicious,bilge2014exposure,antonakakis2011detecting}. Furthermore, whereas ``Standard deviation of the TTL'' has distinct values for benign and malicious domains, it is shown that an intelligent adversary can easily manipulate this feature, leading to a benign classification of malicious domains.

\begin{figure*}[!]
    \hspace*{-1cm}
    \centering
    \includegraphics[width=15cm]{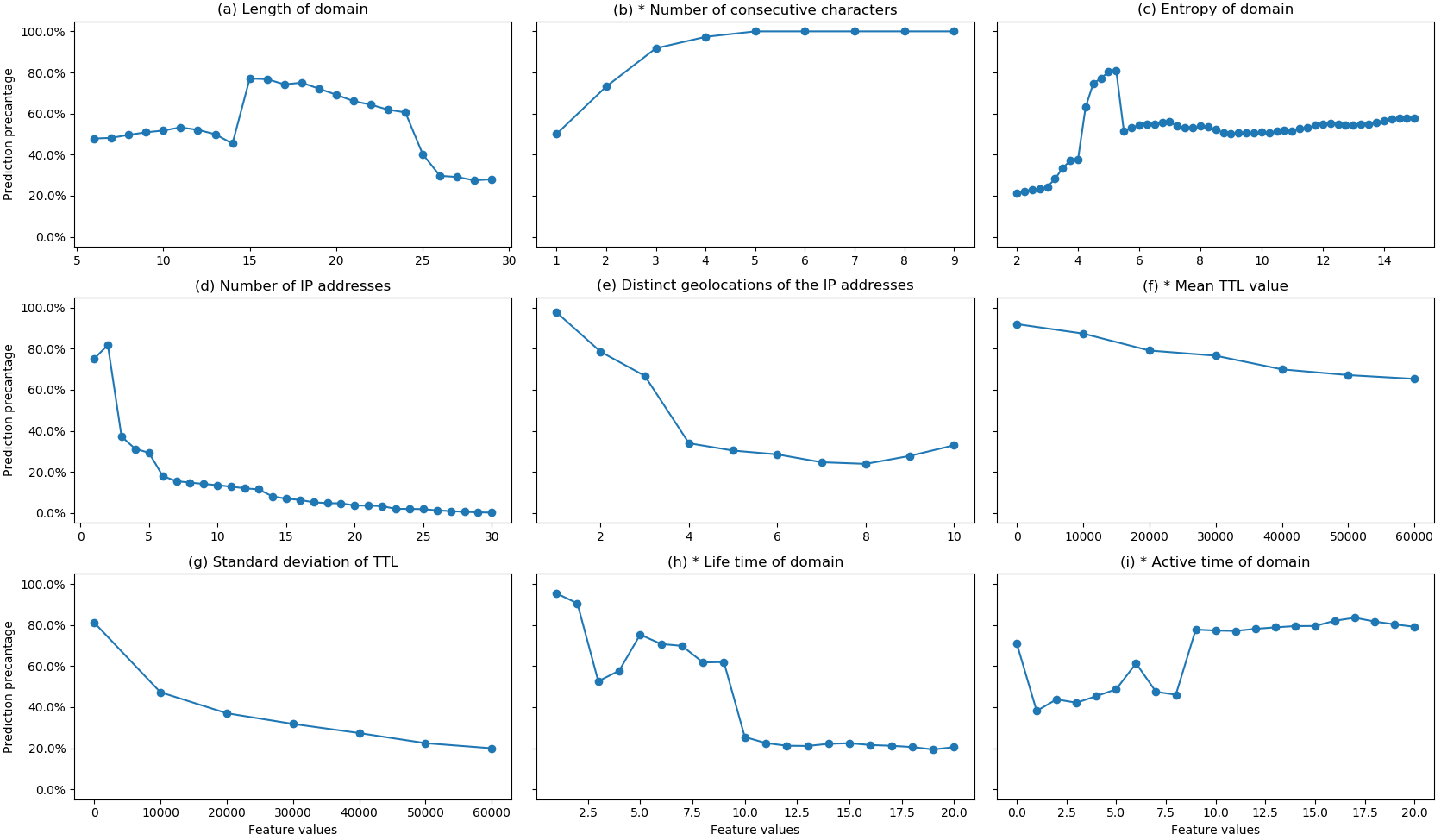}
    \caption{Base feature manipulation graphs (* robust or semi-robust features)}
    \label{fig:basefeaturesmanipulation}
\end{figure*}

\begin{table*}
             \centering
             \addtolength{\leftskip} {-0.6cm}
             \begin{tabular}[t]{|p{6cm}|p{3.4cm}|p{3.4cm}|}
                 \hline
                 \textbf{Feature} & \textbf{Benign mean (std)} & \textbf{Malicious mean (std)} \\
                 \hline\hline
                            Length of domain                          & 14.38 (4.06) & 15.54 (4.09) \\
                            \hline
                            Number of consecutive characters          & 1.29 (0.46) & 1.46 (0.5) \\
                            \hline
                            Entropy of the domain                     & 4.85 (1.18) & 5.16 (1.34) \\
                            \hline
                            Number of IP addresses                    & 2.09 (1.25) & 1.94 (0.94) \\
                            \hline
                            Distinct geolocations of the IP addresses & 1.00 (0.17) & 1.02 (0.31) \\
                            \hline
                            Mean TTL value                           
& 7,578.13 (17,781.47) & 8,039.92 (15,466.29) \\
                            \hline
                            Standard deviation of the TTL             & 2,971.65 (8,777.26) & 2,531.38 (7,456.62) \\
                            \hline
                            Lifetime of domain                      
& 10.98 (7.46) & 6.75 (5.77) \\
                            \hline
                            Active time of domain                     & 8.40 (6.79) & 4.64 (5.66) \\
                            \hline
             \end{tabular}
             \caption{Classic features - statistical properties}
             \label{table:base_features}
\end{table*}
In order to understand the malicious abilities of an adversary, the base features were manipulated over a wide range of possible values, one feature at a time.\footnote{Each feature was evaluated over all possible values for that feature.} For each feature only the range of possible values were taken into account. This analysis considers an intelligent adversary with black-box access to the model (i.e. a set of features or output for a given input). The robustness analysis is based on an ANN model that classifies the manipulated samples, where the train set is the empirically crawled data, and the test set includes the manipulated malicious samples. Figure~\ref{fig:basefeaturesmanipulation} depicts the possible adversary manipulations over any of the features. 
The evaluation metric, the prediction percentage, was defined as the average detection rate after modification.
\vspace{0.3cm}

The well-known features were divided into three groups: robust features, robust features that seemed non-robust (defined as semi-robust), and non-robust features. Next, it can be seen how an attacker can manipulate the classifier for each feature and define its robustness:
\begin{enumerate}
    \item \textbf{''Length of domain'':} an adversary can easily purchase a short or long domain to result in a benign classification for a malicious domain; hence this feature was classified as non-robust.
    \item \textbf{''Number of consecutive characters'':} surprisingly, as depicted in Figure~\ref{fig:basefeaturesmanipulation}, manipulating the ''Number of consecutive characters'' feature can significantly lower the prediction percentage (e.g., move from three consecutive characters to one or two). Nevertheless, as depicted in Table~\ref{table:base_features}, on average, there were 1.46 consecutive characters in malicious domains. Therefore, manipulating this feature is not enough to break the model, and it is considered to be a robust feature.
    \item \textbf{''Entropy of the domain'':} in order to manipulate the ''Entropy of the domain'' feature as benign domain entropy, the adversary can create a domain name with entropy $<$ 4. Take, for example, the domain “ddcd.cc” which is available for purchase. The entropy for this domain is  $3.54$. This value falls precisely in the entropy area of the benign domains defined by the trained model. This example breaks the model and causes a malicious domain to look like a benign URL. Hence, this feature was classified as non-robust.
    \item \textbf{''Number of IP addresses'':} note that an adversary can add many A records to the DNS zone file of its domain to imitate a benign domain. Thus, to manipulate the number of IP addresses, an intelligent adversary only needs to have several different IP addresses and add them to the zone file. This fact classifies this feature as non-robust.
    \item \textbf{''Distinct Geolocations of the IP addresses'':} in order to be able to break the model with the ''Distinct Geolocations of the IP addresses'' feature, the adversary needs to use several IP addresses from different geolocations. If the adversary can determine how many different countries are sufficient to mimic the number of distinct countries of benign domains, he will be able to append this number of IP addresses (a different IP address from each geo-location) to the DNS zone file. Thus, this feature was also classified as non-robust.
    \textbf{''Mean TTL value''} and \textbf{''Standard deviation of the TTL'':} there is a clear correlation between the ''Mean TTL value'' and the ''Standard deviation of the TTL'' features since the value manipulated by the adversary is the TTL itself. Thus, it makes no difference if the adversary cannot manipulate the ''Mean TTL value'' feature if the model uses both. In order to robustify the more,  it is better to use the ''Mean TTL value'' feature without the ''Standard deviation of the TTL'' one. Solely in terms of the ''Mean TTL value'' feature, Figure~\ref{fig:basefeaturesmanipulation} shows that manipulation will not result in a false classification since the prediction percentage does not drop dramatically, even when this feature is drastically manipulated. Therefore this feature is considered to be robust.
    
    An adversary can set the DNS TTL values to [0,120000] (according to the RFC 2181~\cite{rfc2181} the TTL value range is from 0 to $2^{31}-1$). Figure~\ref{fig:basefeaturesmanipulation} shows that even manipulating the value of this feature to 60000 will break the model and cause a malicious domain to be wrongly classified as a benign URL. Therefore the ''Standard deviation of the TTL'' cannot be considered a robust feature.  
    
    \item \textbf{''Lifetime of domain''}: As for the lifetime of domains, based on~\cite{shi2017malicious} we know that a benign domain's lifetime is typically much longer than a malicious domain's lifetime. In order to break the model by manipulating the ''Lifetime of domain'' feature, the adversary must buy an old domain that is available on the market. Even though it is possible to buy an appropriate domain, it will take time to find one, and it will be expensive. Hence we considered this to be a semi-robust feature.
    \item \textbf{''Active time of domain''}: Similar to the previous feature, in order to break ''Active time of domain'', an adversary must find a domain with a particular active time (Figure~\ref{fig:basefeaturesmanipulation}), which is much more tricky. It is hard, expensive, and possibly unfeasible. Therefore this was considered to be a semi-robust feature.
\end{enumerate}

Based on the analysis above, the robust features from Table~\ref{table:base_features} were selected, and the non-robust ones were dropped. Using that subset the model was trained and an accuracy of 95.71\% with an F1-score of 88.78\% was achieved can be improved. Therefore, we extended our analysis and searched for new features that would meet the robustness requirements to build a robust model with a higher F1-score, ending up with a model that results in an accuracy of 99.36\% and F1-score of 98.42\%.

\subsection{Novel Features}
\label{sec:new_fea}
This section presents four novel features which are both robust, and improve the model's efficiency. 

\begin{figure}
        \centering
            \includegraphics[width=\linewidth, height=15 cm]{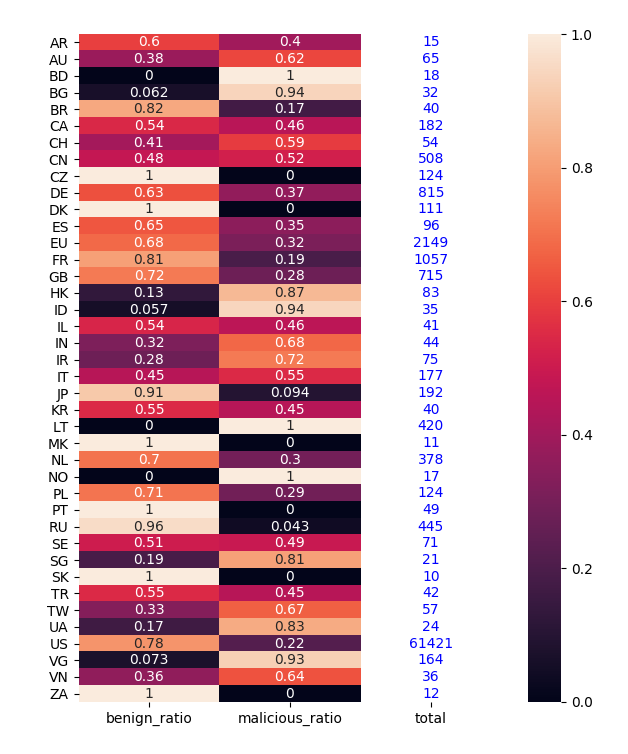}
           \caption{Communication Countries Ratio}
            \label{fig:countries_10}
    \end{figure}
\begin{figure}
        \centering            \includegraphics[width=\linewidth, height=15 cm]{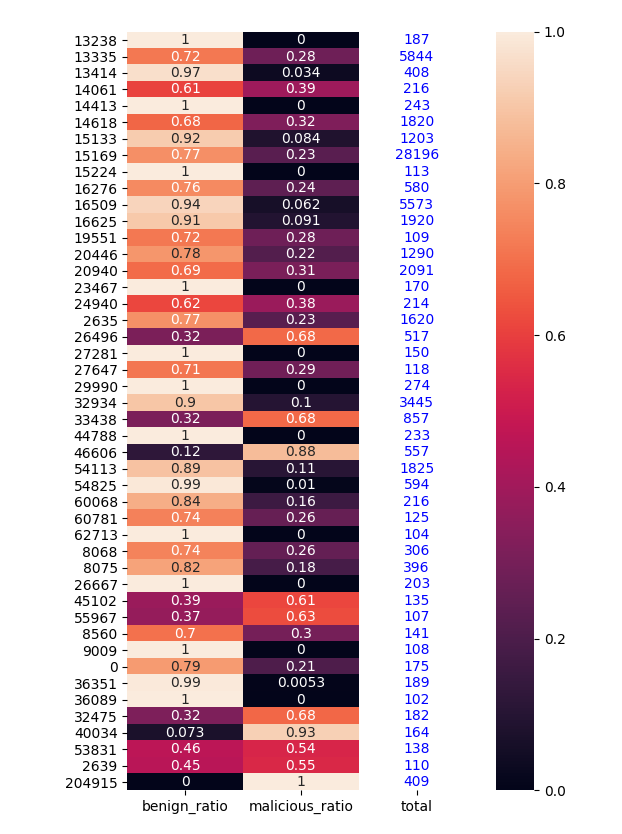}
            \caption{Communication ASNs Ratio}
            \label{fig:asns_100}
    \end{figure}

As stated above, mimicking benign URLs in order to bypass is a mammoth problem. The aim of the research was to validate that manipulating the features in order to result in misclassification of malicious instances will require a disproportionate effort that will deter the attacker from doing so. The four novel features were designed according to this paradigm based on two communication information properties, passive DNS changes, and the remaining time of SSL certificate. For each IP, Urlscan extract the geo-location, which in turn is appended to a communication country list. Similarly, the communication ASNs is a list of ASNs, that was extracted using Urlscan, each IP address, and appended the ASNs list. Using the URL dataset and the Urlscan service, benign-malicious ratio tables for communication countries and for communication ASNs (Figures~\ref{fig:countries_10},\ref{fig:asns_100}) were created.
The ratio tables were calculated for each element \textit{E} (country - for the communication countries ratio table, or ASN - for the communication ASNs ratio table). Each table represents the probability that a URL associated with a country (ASN) is malicious. In order to extract those probabilities, the number of malicious URLs associated with \textit{E} was divided by the total URLs associated with \textit{E}. Initially, due to the heterogeneity of the dataset (i.e. there exist some elements that appear only a few times), the ratio tables were seen to be biased. To overcome this challenge, an initial threshold was set as an insertion criteria which is later detailed in Algorithm~\ref{rank_communication} (CRA).

The following is a detailed summary of the novel features:
\begin{itemize}
    \item \textbf{Communication Countries Rank (CCR)}:
   \begin{multline}
            CCR = \\ \operatorname{CRA}(\textrm{the communication countries list of } URL_{(i)})
    \end{multline}
    This feature looks at the communication countries with respect to the communication IPs, and uses the countries ratio table to rank a specific URL. 
    
    \item \textbf{Communication ASNs Rank (CAR)}:
     \begin{multline}
            CAR = \\ \operatorname{CRA}(\textrm{the communication ASNs list of } URL_{(i)})
    \end{multline}
    Similarly, this feature analyzes the communication ASNs with respect to the communication IPs, and uses the ASNs ratio table to rank a specific URL. 
    While there is some correlation between the ASNs and the countries, the second feature examines each AS (usually ISPs or large companies) within each country to gain a wider perspective.
    
    \item \textbf{Number of passive DNS changes}:
    \begin{multline}
            PDNS = \\ \operatorname{count}(\textrm{list of passive DNS records of } Domain_{(i)})
    \end{multline}
    When inspecting the passive DNS records, benign domains emerged as having much larger DNS changes that the sensors (of the company that collects the DNS records) could identify, unlike malicious domains (i.e. 26.4 vs. 8.01, as reported in Table~\ref{table:extended_features}).
    
    For the ”Number of passive DNS changes” the number of DNS records changes were counted, which is somewhat similar to other features described in ~\cite{torabi2018detecting,antonakakis2010building}. Still, these features require much more elaborated information which is not publicly available. On the other hand, this feature can be extracted from passive DNS records obtained from VirusTotal, which are scarce (in terms of record types).
    
    \item \textbf{Remaining time of SSL certificate}:
    \begin{multline}
            SSL = Certificate_{Valid}  \cdot \\  \cdot(Certificate_{Expiration}-Certificate_{Updated})
    \end{multline}
    When installing an SSL certificate, there is a validation process conducted by a Certificate Authority (CA). Depending on the type of certificate, the CA verifies the organization's identity before issuing the certificate. When analyzing our data it was noted that most of the malicious domains do not use valid SSL certificates and those that do only use one for a short period. Therefore, this feature was engineered which represents the time the SSL certificate remains valid. 

    For the ”Remaining time of SSL certificate”, in contrast to a binary feature version used by~\cite{ranganayakulu2013detecting}, this feature extends the scope and represents both the existence of an SSL certificate and the remaining time until the SSL certificate expires.
\end{itemize}

\begin{algorithm}[H]
    \caption{Communication Rank}
    \label{rank_communication}
    \textbf{Input:} URL, Threshold, Type\\
    \textbf{Output:} Rank (CCR or CAR)
    \begin{algorithmic}[1]
    \IF{Type = Countries}
        \STATE ItemsList = communication countries list of the URL
    \ELSE
    	\STATE{ItemsList = ASNs list of the URL}
    \ENDIF
    \STATE $Rank = 0$
    \FOR{$Item$ in $ItemsList$}
        \STATE $Ratio=0.75$ \COMMENT{Init value}
        \STATE $Total\_norm=1$ \COMMENT{Init value}
        \IF{$TotalOccurrences(Item)>=Threshold$}
            \STATE $Total\_norm=Normalize(Item)$
            \STATE $Ratio=BenignRatio(Item)$
        \ENDIF
        \STATE $Rank+=(\log_{0.5}(Ratio+\epsilon)/Total\_norm)$
    \ENDFOR
    \end{algorithmic}
    \end{algorithm}

    The CRA (Algorithm~\ref{rank_communication}) gets a URL as an input and returns its country communication rate or the ASN communication rate (based on the type in the input of the algorithm).
    
    For each item (i.e., country or ASN), first the algorithm initialized the value of the ratio variable to 0.75 (according to~\cite{webrootsite}, 25\% of all URLs in 2017 were malicious, suspicious or moderately risky) and the normalized total occurrences (Total\_norm) of an item  to be 1.
    Next, in Step 9, if the total number of occurrences of an item was $\geq$ to the threshold, the algorithm replaced the ratio and normalized occurrences to the correct values according to the ratio tables given in Figures~\ref{fig:countries_10} and ~\ref{fig:asns_100}. Finally, the algorithm sums the rank with a log base 0.5 of the ratio (+ some epsilon) and divide this value by the normalized total occurrences. 
    

\begin{table}
	\centering
	\begin{tabular}{|p{6cm}|p{3cm}|p{3cm}|}\hline
		\centering
	    \textbf{Feature} & \textbf{Benign mean (std)} & \textbf{Malicious mean (std)} \\\hline
	    \centering
		Communication Countries Rank (CCR) & 31.31 (91.16) & 59.40 (215.15) \\ 
		\hline
		\centering
		Communication ASNs Rank (CAR)      & 935.59 (12,258.99) & 12,979.38 (46,384.86) \\
		\hline
		\centering
		Number of passive DNS changes     & 26.40 (111.99) & 8.01 (16.63) \\
		\hline
		\centering
		Remaining time of SSL certificate & 1.547E7 (2.304E7) & 4.365E6 (1.545E7) \\
		\hline
	\end{tabular}
    \caption{Novel features - statistical properties}
	\label{table:extended_features}
\end{table}


\begin{figure*}
	\centering
	\includegraphics[width=14cm]{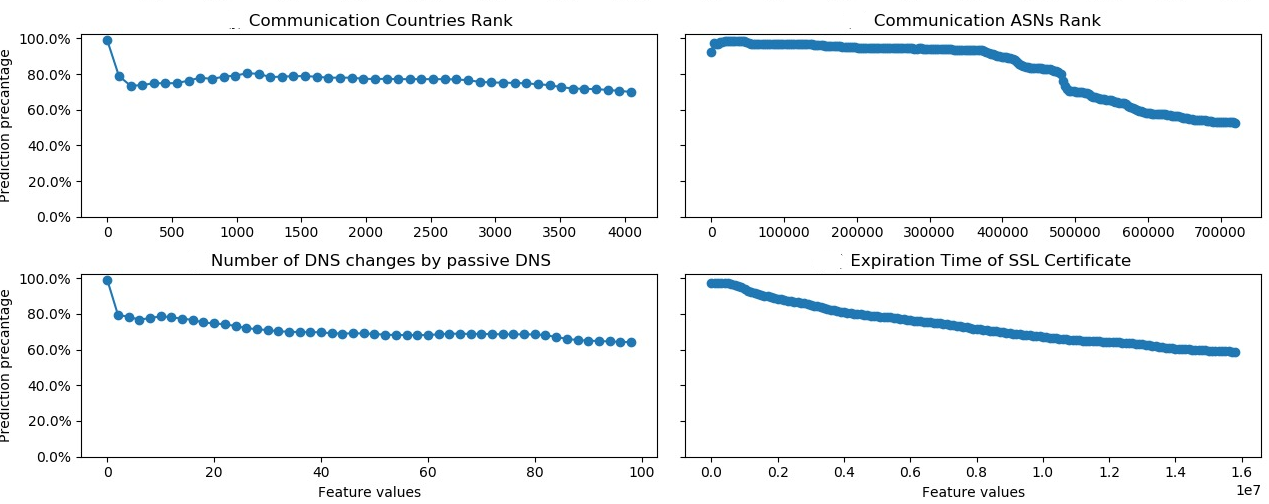}
	\caption{Novel robust feature manipulation graphs}
	\label{fig:ourfeaturesmanipulation}
\end{figure*}

Figure~\ref{fig:ourfeaturesmanipulation} depicts the prediction percentage as a function of changing the novel features values for each feature in Table~\ref{table:extended_features}, Similar to the analysis in Section~\ref{sec:robust}. This evaluation proves that manipulating the values of the novel features does not break the robust model (i.e., the prediction percentage remains steady). While one may be concerned by the negative correlation between ''Remaining time of SSL certificate'' feature and the prediction percentage, note that the average value for malicious domains is three times higher than the benign ones. While theoretically the adversary can lower this value, the implications of such action are acquiring (or use some free solution) an SSL certificate. Since there is a validation process, this process will cause the adversary to lose its anonymity and be identified.







\section{Empirical Analysis and Evaluation}
\label{Chapter5}
This section describes the testbed used for the evaluation of models based on the types of features (both robust and not). General settings are provided for each of the models (e.g. the division of the data into training and test set), as are the parameters used to configure each of the models, followed by the efficiency of each model.~\footnote{Our code is publicly available at~\url{https://github.com/nitayhas/robust-malicious-url-detection}} 

\subsection{Experimental Design}\label{sec:experimental_design}
Apart from intelligently choosing the model parameters, one should verify that the data used for the learning phase accurately represents the real-world distribution of domain malware. Hence, the dataset was constructed such that 75\%  were benign domains, and the remaining 25\% were malicious domains (\textasciitilde5,000 benign URLs and  \textasciitilde1,350 malicious domains respectively)~\cite{webrootsite}.

There are many ways to define the efficiency of a model. To account for most of them, a broad set of metrics was extracted including accuracy, recall, F1-score, and training time. Note that for each model, the dataset was split into train and test sets where 75\% of the data (both benign and malicious) was randomly assigned to the train test, and the remaining domains are assigned to the test set.

The evaluation step measured the efficiency of the different models while varying the robustness of the features included in the model. Specifically, four different models (i.e. Logistic Regression, SVM, ELM, and ANN) were trained using the following feature sets:  
\begin{itemize}
    \item Base (\textit{B}) -  The set of commonly used features in previous works (see  Table~\ref{table:base_features} for more details).
    \item Base Robust (\textit{BR}) - the subset of robust base features  (marked with a * in Figure~\ref{fig:basefeaturesmanipulation}).
    \item ''TCP'' (\textit{TCP}) - The four novel features: Time of SSL certificate, Communication ranks (CCR and CAR) and PassiveDNS changes (see Table~\ref{table:extended_features}).
    \item Base Robust + ''TCP''(\textit{BRTCP}) - the union of \textit{BR} and \textit{TCP}, the robust subset of all features.
    \item Base + ''TCP'' (\textit{BTCP}) - the union of \textit{B} and \textit{TCP}.
\end{itemize}

\begin{figure}[ht]
    \begin{subfigure}{0.5\textwidth}
      \centering
      \includegraphics[width=0.8\linewidth]{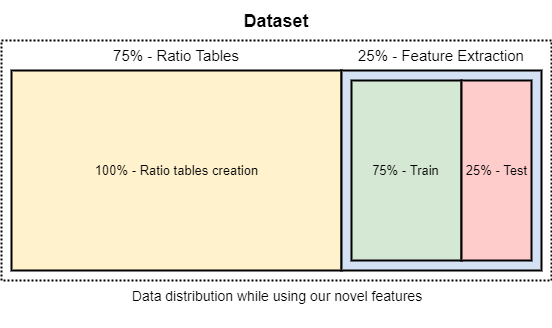}  
      \label{fig:sub-first}
      \caption{Data distribution while using our novel features}
    \end{subfigure}
    \begin{subfigure}{0.5\textwidth}
      \centering
      \includegraphics[width=0.8\linewidth]{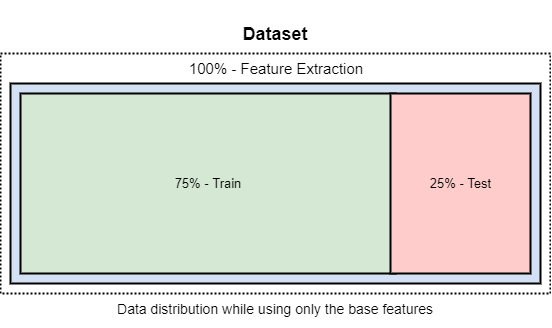}  
      \label{fig:sub-second}
      \caption{Data distribution while using only the base features}
    \end{subfigure}
\caption{Data Distribution}
\label{fig:models_accuracy}
\end{figure}

Recall that feature sets (i.e. TCP, BRTCP, and BTCP) which are based on the novel features (i.e. CCR and CAR) , require communication ratio tables. Hence, to keep the models unbiased (i.e. omit the data that was used for generating the ratio tables from the learning process), it was necessary allocate some of the data to creation of the ratio tables. Due to the variety of countries and ASNs, it was decided to dedicate the majority of the data to this process. For the evaluation, the dataset was split into two parts: when using the novel features 75\% of the data was used to create the ratio tables and the remaining 25\% of the data was used to extract the features, train and test for the models, as can be seen in Figure \ref{fig:models_accuracy}a. In the case of using the well-known features, 100\% of the data was used for the feature extraction as can be seen in Figure \ref{fig:models_accuracy}b. Evaluations for the opposite split ratio (i.e. 25/75) were also conducted, this time allocating most of the data to the learning phase. The findings showed that the trend between these two ratios was similar (as presented in Section~\ref{sec:experimental_results}). Both ratios returned high results but, for most of the models, the 75/25 ratio returned a higher F1-Score and higher Recall measures (e.g. LR, SVM, and ANN for each feature set). For the 25/75 distribution, the F1-Score and the Recall measures were only higher for the ELM (around 1\%-2\% higher). Therefore, while both results are reported in the end of this section, WLOG, the main focus of the discussion is on the 75/25 distribution.





\subsection{Models and Parameters}
\label{sec:experimental_results}
Four commonly used classification models are analyzed: Logistic Regression (LR), Support Vector Machines (SVM), Extreme Learning Machine (ELM), and Artificial Neural Networks (ANN). All the models were trained and evaluated on a Dell XPS 8920 computer, Windows 10 64Bit OS with 3.60GHz Intel Core i7-7700 CPU, 16GB of RAM, and NVIDIA GeForce GTX 1060 6GB. 

In the following paragraphs, for each model, the hyperparameters used for the evaluation are first described, followed by the empirical, experimental results (which sums up several test results using different random train-test sets), and a short discussion of the findings and their implications.

\subsubsection{Logistic Regression}
As a baseline for the evaluation process, and before using the nonlinear models, the LR classification model was used. The LR model with the five feature sets was trained and the hyperparameters were tuned to maximize the model's performance.

\textbf{Hyperparameters:} Polynomial degree: 3;K-Fold Cross-Validation k=10; Solver: L-BFGS~\cite{liu1989}.

Table~\ref{table:lr_results} shows that the different feature sets resulted in similar accuracy rates. However, the accuracy rate measures how well the model predicts (i.e. TP+TN) with respect to all the predictions (i.e. TP+TN+FP+FN). Thus given the unbalanced dataset (75\% of the dataset are benign and 25\% are malicious domains), \textasciitilde90\% accuracy is not necessarily a sufficient result in terms of malware detection. For example, the \textit{TCP} feature set has high accuracy but in contrast a very poor F1-Score, due to the high Precision rate and poor Recall rate (which represents the ratio of malicious instances detected). 
As the recall is low for all features sets, this work demonstrates that accuracy rate is not a good measure in this domain; therefore, it was decided to focus on the F1-score measure, which is the harmonic mean of the precision and the recall measures. Next, it was decided to use the SVM model with an RBF kernel as a nonlinear model.

\subsubsection{Support Vector Machine (SVM)}
\textbf{Hyperparameters:} Polynomial degree: 3; K-Fold Cross-Validation k=10; $\gamma$=2; Kernel: RBF~\cite{park1991universal}.

Compared to the results of the LR model (Table~\ref{table:lr_results}), the results of the SVM model (Table~\ref{table:svm_results}) show a significant improvement in the recall and F1-score measures; e.g. for \textit{Base}, the recall and the F1-score measures were both above 90\%. One could be concerned by the fact that the model trained on the \textit{Base} feature set resulted in a higher recall (and F1-score) compare to the one trained on the \textit{Robust Base} feature set. However, it should be noted that the \textit{Robust Base} feature set is robust to adversarial manipulation and uses less than half of the features provided in the training phase with the \textit{Base} feature set. This discussion also applies to the \textit{BRTCP} and \textit{BTCP} feature sets. Another advantage of including the novel features is the fact that models converge much faster. 

The results are based on analyzing a non-manipulated dataset. As stated above, the \textit{Base} feature set includes some non-robust features. Hence, an intelligent adversary can manipulate the values of these features, resulting in a wrong classification of malicious instances (up to the extreme of a $ 0\% $ recall). However, an intelligent adversary should invest much more effort for a model that was trained using the \textit{Robust Base} or \textit{TCP} features, since each of them was specifically chosen to avoid such manipulations. 
In order to find models that were also efficient on the non-manipulated dataset, the two sophisticated models were examined in the analysis, the ELM model as presented in~\cite{shi2017malicious} and the ANN model.
\begin{table}
    \centering
        \begin{tabular}{|c|c|c|c|}
        \hline
        \textbf{Feature set} & \textbf{Accuracy} & \textbf{Recall} & \textbf{F1-Score} \\
        \hline\hline
        \textit{Base} & 89.99\% & 38.82\% & 53.21\%\\
        \hline
        \textit{Robust Base} & 88.33\% & 38.87\% & 49.42\%\\
        \hline
        \textit{TCP}  & 86.20\% & 8.30\% & 14.99\%\\
        \hline
        \textit{BRTCP} & 88.82\% & 52.46\% & 65.57\%\\
       \hline
       \textit{BTCP} & 92.86\% & 64.14\% & 72.48\%\\
        \hline
    \end{tabular}
    \caption{Model performance - Logistic Regression (in \%)}
    \label{table:lr_results}
\end{table}
\begin{table}
    \centering
    \begin{tabular}{|c|c|c|c|}
        \hline
        \textbf{Feature set} & \textbf{Accuracy} & \textbf{Recall} & \textbf{F1-Score}\\
        \hline\hline
        \textit{Base} & 96.49\% & 91.20\% & 91.36\%\\
        \hline
        \textit{Robust Base} & 90.14\% & 56.51\% & 69.93\%\\
        \hline
        \textit{TCP}  & 83.10\% & 60.21\% & 54.21\%\\
        \hline
        \textit{BRTCP} & 96.78\% & 91.37\% & 92.02\%\\
        \hline
        \textit{BTCP} & 97.95\% & 90.73\% & 92.83\%\\
        \hline
    \end{tabular}
    \caption{Model performance - SVM (in \%)}
    \label{table:svm_results}
\end{table}

\subsubsection{ELM}
\textbf{Hyperparameters}: One input layer, one hidden layer, and one output layer. Activation function: first layer - ReLU~\cite{nair2010rectified}; hidden layer - Sigmoid. K-Fold Cross-Validation k=10 \cite{shi2017malicious}.
Overall, the ELM model resulted in high accuracy and higher Recall rates compared to Table~\ref{table:lr_results}, for any feature set. When compared to the SVM models, the \textit{Base} model resulted in lower recall (though a higher F1-score was achieved with the ELM model). On the other hand, the \textit{Robust Base} resulted in a higher recall in the ELM model compared to the SVM one. Even though the \textit{Robust Base} feature set had a low dimensional space, the three rates (i.e. Accuracy, Recall, and F1-score) were higher than those of the \textit{Base} feature set. Moving to the sets that include the novel features increased these metrics, while improving the robustness of the model at the same time. 

\begin{table}
    \centering
    \begin{tabular}{|c|c|c|c|}
        \hline
        \textbf{Feature set} & \textbf{Accuracy} & \textbf{Recall} & \textbf{F1-Score}\\
        \hline\hline
        \textit{Base} & 98.17\% & 88.81\% & 92.92\%\\
        \hline
        \textit{Robust Base} & 98.83\% & 92.24\% & 95.81\%\\
        \hline
        \textit{TCP}  & 98.88\% & 94.64\% & 96.84\%\\
        \hline
        \textit{BRTCP} & 98.86\% & 95.82\% & 97.07\%\\
        \hline
        \textit{BTCP} & 98.19\% & 93.09\% & 95.34\%\\
        \hline
    \end{tabular}
    \caption{Model performance - ELM (in \%)}
    \label{table:elm_results}
\end{table}

\subsubsection{ANN}
\textbf{Hyperparameters:} One input layer, three hidden layers, and one output layer. Activation function: first layer - ReLU; first hidden layer - RELU; Second hidden layer - LeakyReLU; Third hidden layer - Sigmoid. Batch size -150, learning rate of 0.01; Solver: Adam~\cite{kingma2014adam} with $\beta_1$  = 0.9 and $\beta_2 = 0.999$. K-Fold Cross-Validation k=10.
Similarl to the ELM results, the ANN results show high performance on all feature sets. 
For the ``basic'' feature sets (i.e. \textit{Base} and \textit{Robust Base}) the ELM models resulted in higher recall and F1-score. Still, the main focus was in the \textit{BTCP} feature set and more specifically in the \textit{BRTCP} variant, and for those feature sets, the ANN models resulted in higher recall and F1-score. 
\begin{table}
    \centering
    \begin{tabular}{|c|c|c|c|}
        \hline
        \textbf{Feature set} & \textbf{Accuracy} & \textbf{Recall} & \textbf{F1-Score}\\
        \hline\hline
        \textit{Base} & 97.20\% & 88.03\% & 90.23\%\\
        \hline
        \textit{Robust Base} & 95.71\% & 83.63\% & 88.78\%\\
        \hline
        \textit{TCP}  & 98.03\% & 96.83\% & 95.24\%\\
        \hline
        \textit{BRTCP} & 99.36\% & 98.77\% & 98.42\%\\
       \hline
        \textit{BTCP} & 99.82\% & 99.47\% & 99.56\%\\
        \hline
    \end{tabular}
    \caption{Model performance - ANN (in \%)}
    \label{table:ann_results}
\end{table}

\subsubsection{Discussion}
\begin{figure}
	\centering
	\includegraphics[width=0.5\textwidth]{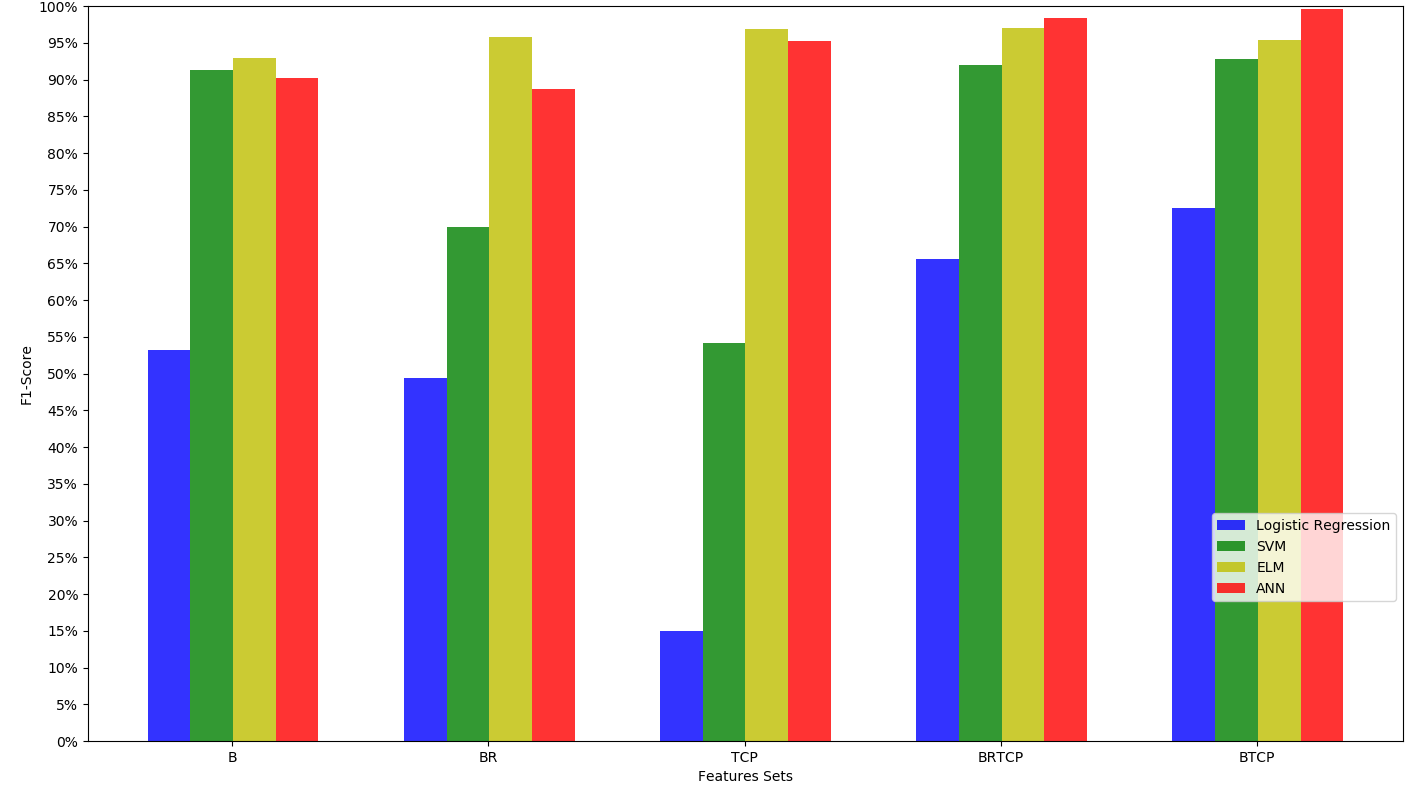}
	\caption{The F1-Score by feature sets and models}
	\label{fig:models_f1}
\end{figure}

This analysis conclude with Figure~\ref{fig:models_f1} and Tables \ref{table:full_results_a}, \ref{table:full_results_b} and \ref{table:full_results_c} that summarize the evaluation. In particular, Figure~\ref{fig:models_f1} shows the overall results and presents the F1-scores of the feature sets for all the models.

	\begin{table}
    \centering
    \begin{tabular}{|p{2cm}|p{4cm}|p{4cm}|}
   \hline
    \diaghead{\theadfont Diag ColumnmnHead II}%
{Model}{Feature set} & Base & Base Robust \\ \hline
    Logistic Regression & 
    \begin{minipage}{3.4cm}
    \vspace{0.2cm}
    Accuracy: 0.90\\
    Precision: 0.84\\
    Recall: 0.39\\
    F1-score: 0.53\\
    Loss: 3.45\\
    AUC: 0.85
    \vspace{0.2cm}
    \end{minipage}&
    \begin{minipage}{3.4cm}
    \vspace{0.2cm}
    Accuracy: 0.88\\
    Precision: 0.68\\
    Recall: 0.39\\
    F1-score: 0.49\\
    Loss: 4.03\\
    AUC: 0.81
    \vspace{0.2cm}
    \end{minipage}
    \\ \hline
    SVM &
    \begin{minipage}{3.4cm}
    \vspace{0.2cm}
    Accuracy: 0.96\\
    Precision: 0.91\\
    Recall: 0.91\\
    F1-score: 0.91\\
    Loss: 1.20\\
    AUC: 0.96
    \vspace{0.2cm}
    \end{minipage}&
    \begin{minipage}{3.4cm}
    \vspace{0.2cm}
    Accuracy: 0.90\\
    Precision: 0.91\\
    Recall: 0.56\\
    F1-score: 0.69\\
    Loss: 3.40\\
    AUC: 0.92
    \vspace{0.2cm}
    \end{minipage}
    \\ \hline
    ELM & 
    \begin{minipage}{3.4cm}
    \vspace{0.2cm}
    Accuracy: 0.98\\
    Precision: 0.98\\
    Recall: 0.88\\
    F1-score: 0.92\\
    Loss: 0.63\\
    AUC: 0.99
    \vspace{0.2cm}
    \end{minipage}&
    \begin{minipage}{3.4cm}
    \vspace{0.2cm}
    Accuracy: 0.98\\
    Precision: 0.99\\
    Recall: 0.92\\
    F1-score: 0.95\\
    Loss: 0.40\\
    AUC: 0.99
    \vspace{0.2cm}
    \end{minipage}
    \\ \hline
    ANN &
    \begin{minipage}{3.4cm}
    \vspace{0.2cm}
    Accuracy: 0.97\\
    Precision: 0.92\\
    Recall: 0.88\\
    F1-score: 0.90\\
    Loss: 0.44\\
    AUC: 0.98
    \vspace{0.2cm}
    \end{minipage}&
    \begin{minipage}{3.4cm}
    \vspace{0.2cm}
    Accuracy: 0.95\\
    Precision: 0.94\\
    Recall: 0.83\\
    F1-score: 0.88\\
    Loss: 0.68\\
    AUC: 0.97
    \vspace{0.2cm}
    \end{minipage}
    \\ \hline
    \end{tabular}
    \caption{Raw data results for the models that trained with the base features (100\% of the dataset records)}
    \label{table:full_results_a}
\end{table}
Tables \ref{table:full_results_a}-\ref{table:full_results_c} depicts all the evaluation matrices (i.e. accuracy, precision, recall, F1-score, loss, and AUC). While Table \ref{table:full_results_b} depicts the models' performance where 75\% of the data was used to construct the ratio table, Table \ref{table:full_results_c} depicts the performance where the minority of data (i.e. 25\%) was used to construct the ratio tables.
Looking at the resultant recall and F1-score, it seems that models analyzed in Table~\ref{table:full_results_b}, in which most of the data is dedicated to the ratio tables (i.e. to the TCP features), resulted in higher performance, even though these models had one third of the data available to the models evaluated in Table~\ref{table:full_results_c}. At the same time, it is important to note that the best model resulted by using the Base Robust + TCP features, to train the ELM model and dedicating 25\% of the data to the feature engineering phase, and the reminder 75\% to the learning process.

All the results provided above are based on clean data (i.e. with no adversarial manipulation). Naturally, given an adversarial model where the attacker can manipulate the values of features, models which are based on the \textit{Robust Base} or \textit{TCP} feature sets will dominate models that are trained using the \textit{Base} dataset. Thus, by showing that the \textit{Robust Base} feature set does not dramatically decrease the performance of the classifier using clean data, and that adding the novel feature improves the model's performance as well as its robustness, the conclusion is that malicious domain classifiers should use this feature set for robust malicious domain detection.



\begin{table}
    \centering
 \centering
    \begin{tabular}{|p{3.5cm}|p{3cm}|p{3cm}|p{3cm}|}
   \hline
 Model/Feature set & TCP & Base Robust + TCP & Base + TCP \\ \hline
    Logistic Regression & 
    \begin{minipage}{3.4cm}
    \vspace{0.2cm}
    Accuracy: 0.86\\
    Precision: 0.77\\
    Recall: 0.08\\
    F1-score: 0.15\\
    Loss: 4.76\\
    AUC: 0.79
    \vspace{0.2cm}
    \end{minipage}&
    \begin{minipage}{3.4cm}
    \vspace{0.2cm}
    Accuracy: 0.88\\
    Precision: 0.87\\
    Recall: 0.52\\
    F1-score: 0.65\\
    Loss: 3.86\\
    AUC: 0.93
    \vspace{0.2cm}
    \end{minipage}&
    \begin{minipage}{3.4cm}
    \vspace{0.2cm}
    Accuracy: 0.92\\
    Precision: 0.83\\
    Recall: 0.64\\
    F1-score: 0.72\\
    Loss: 2.46\\
    AUC: 0.94
    \vspace{0.2cm}
    \end{minipage}
    \\ \hline
    SVM &
    \begin{minipage}{3.4cm}
    \vspace{0.2cm}
    Accuracy: 0.83\\
    Precision: 0.60\\
    Recall: 0.49\\
    F1-score: 0.54\\
    Loss: 5.83\\
    AUC: 0.80
    \vspace{0.2cm}
    \end{minipage}&
    \begin{minipage}{3.4cm}
    \vspace{0.2cm}
    Accuracy: 0.96\\
    Precision: 0.92\\
    Recall: 0.91\\
    F1-score: 0.92\\
    Loss: 1.11\\
    AUC: 0.98
    \vspace{0.2cm}
    \end{minipage}&
    \begin{minipage}{3.4cm}
    \vspace{0.2cm}
    Accuracy: 0.97\\
    Precision: 0.95\\
    Recall: 0.90\\
    F1-score: 0.92\\
    Loss: 0.70\\
    AUC: 0.98
    \vspace{0.2cm}
    \end{minipage}
    \\ \hline
    ELM & 
    \begin{minipage}{3.4cm}
    \vspace{0.2cm}
    Accuracy: 0.98\\
    Precision: 0.99\\
    Recall: 0.94\\
    F1-score: 0.96\\
    Loss: 0.38\\
    AUC: 0.99
    \vspace{0.2cm}
    \end{minipage}&
    \begin{minipage}{3.4cm}
    \vspace{0.2cm}
    Accuracy: 0.98\\
    Precision: 0.98\\
    Recall: 0.95\\
    F1-score: 0.97\\
    Loss: 0.39\\
    AUC: 0.99
    \vspace{0.2cm}
    \end{minipage}&
    \begin{minipage}{3.4cm}
    \vspace{0.2cm}
    Accuracy: 0.98\\
    Precision: 0.97\\
    Recall: 0.93\\
    F1-score: 0.95\\
    Loss: 0.62\\
    AUC: 0.98
    \vspace{0.2cm}
    \end{minipage}
    \\ \hline
    ANN &
    \begin{minipage}{3.4cm}
    \vspace{0.2cm}
    Accuracy: 0.98\\
    Precision: 0.93\\
    Recall: 0.96\\
    F1-score: 0.95\\
    Loss: 0.31\\
    AUC: 0.99
    \vspace{0.2cm}
    \end{minipage}&
    \begin{minipage}{3.4cm}
    \vspace{0.2cm}
    Accuracy: 0.99\\
    Precision: 0.98\\
    Recall: 0.98\\
    F1-score: 0.98\\
    Loss: 0.10\\
    AUC: 0.99
    \vspace{0.2cm}
    \end{minipage}&
    \begin{minipage}{3.4cm}
    \vspace{0.2cm}
    Accuracy: 0.99\\
    Precision: 0.99\\
    Recall: 0.99\\
    F1-score: 0.99\\
    Loss: 0.02\\
    AUC: 0.99
    \vspace{0.2cm}
    \end{minipage}
    \\ \hline
    \end{tabular}
    \caption{Raw data results for the models that trained on the 75/25 dataset distribution, 42,578/10,979 URLs (ratio tables/feature extraction)}
    \label{table:full_results_b}
\end{table}

\begin{table}
    \centering
 \centering
    \begin{tabular}{|p{3.5cm}|p{3cm}|p{3cm}|p{3cm}|}
   \hline
Model/Feature set
     & TCP & Base Robust + TCP & Base + TCP \\ \hline
    Logistic Regression & 
    \begin{minipage}{3.4cm}
    \vspace{0.2cm}
    Accuracy: 0.82\\
    Precision: 0.80\\
    Recall: 0.07\\
    F1-score: 0.13\\
    Loss: 5.94\\
    AUC: 0.70
    \vspace{0.2cm}
    \end{minipage}&
    \begin{minipage}{3.4cm}
    \vspace{0.2cm}
    Accuracy: 0.87\\
    Precision: 0.85\\
    Recall: 0.39\\
    F1-score: 0.53\\
    Loss: 4.24\\
    AUC: 0.89
    \vspace{0.2cm}
    \end{minipage}&
    \begin{minipage}{3.4cm}
    \vspace{0.2cm}
    Accuracy: 0.90\\
    Precision: 0.82\\
    Recall: 0.57\\
    F1-score: 0.68\\
    Loss: 3.42\\
    AUC: 0.93
    \vspace{0.2cm}
    \end{minipage}
    \\ \hline
    SVM &
    \begin{minipage}{3.4cm}
    \vspace{0.2cm}
    Accuracy: 0.82\\
    Precision: 0.89\\
    Recall: 0.07\\
    F1-score: 0.13\\
    Loss: 5.88\\
    AUC: 0.76
    \vspace{0.2cm}
    \end{minipage}&
    \begin{minipage}{3.4cm}
    \vspace{0.2cm}
    Accuracy: 0.91\\
    Precision: 0.81\\
    Recall: 0.65\\
    F1-score: 0.72\\
    Loss: 3.09\\
    AUC: 0.94
    \vspace{0.2cm}
    \end{minipage}&
    \begin{minipage}{3.4cm}
    \vspace{0.2cm}
    Accuracy: 0.96\\
    Precision: 0.91\\
    Recall: 0.87\\
    F1-score: 0.89\\
    Loss: 1.33\\
    AUC: 0.97
    \vspace{0.2cm}
    \end{minipage}
    \\ \hline
    ELM & 
    \begin{minipage}{3.4cm}
    \vspace{0.2cm}
    Accuracy: 0.99\\
    Precision: 0.99\\
    Recall: 0.96\\
    F1-score: 0.97\\
    Loss: 0.25\\
    AUC: 0.99
    \vspace{0.2cm}
    \end{minipage}&
    \begin{minipage}{3.4cm}
    \vspace{0.2cm}
    Accuracy: 0.99\\
    Precision: 0.99\\
    Recall: 0.98\\
    F1-score: 0.98\\
    Loss: 0.16\\
    AUC: 0.99
    \vspace{0.2cm}
    \end{minipage}&
    \begin{minipage}{3.4cm}
    \vspace{0.2cm}
    Accuracy: 0.98\\
    Precision: 0.97\\
    Recall: 0.93\\
    F1-score: 0.95\\
    Loss: 0.54\\
    AUC: 0.99
    \vspace{0.2cm}
    \end{minipage}
    \\ \hline
    ANN &
    \begin{minipage}{3.4cm}
    \vspace{0.2cm}
    Accuracy: 0.94\\
    Precision: 0.99\\
    Recall: 0.72\\
    F1-score: 0.83\\
    Loss: 0.81\\
    AUC: 0.97
    \vspace{0.2cm}
    \end{minipage}&
    \begin{minipage}{3.4cm}
    \vspace{0.2cm}
    Accuracy: 0.98\\
    Precision: 0.99\\
    Recall: 0.90\\
    F1-score: 0.94\\
    Loss: 0.31\\
    AUC: 0.99
    \vspace{0.2cm}
    \end{minipage}&
    \begin{minipage}{3.4cm}
    \vspace{0.2cm}
    Accuracy: 0.98\\
    Precision: 0.97\\
    Recall: 0.95\\
    F1-score: 0.96\\
    Loss: 0.18\\
    AUC: 0.99
    \vspace{0.2cm}
    \end{minipage}
    \\ \hline
    \end{tabular}
    \caption{Raw data results for the models that trained on the 25/75 dataset distribution 10,979/42,578 URLs (ratio tables/feature extraction)}
    \label{table:full_results_c}
\end{table}

\section{Conclusion}
\label{Chapter6}

Numerous attempts have been made to tackle the problem of identifying malicious domains. However, many of them fail to successfully classify malware in realistic environments where an adversary can manipulate the URLs and/or other extracted features. Specifically, this research tackled the case where an attacker has access to the model (i.e. a set of features or output for a given input), and tampers with the domain properties. This tampering has a catastrophic effect on the model’s efficiency. As a countermeasure, an intelligent feature selection procedure was used which is robust to adversarial manipulation as well as inclusion of novel robust features. Feature robustness and model effectiveness were evaluated based on well known machine and deep learning models over a sizeable realistic dataset. 

The evaluation showed that models that are trained using the robust features are more precise in terms of manipulated data while maintaining good results on clean data as well. Clearly, further research is needed to create models that can also classify malicious domains into malicious attack types. Another promising direction would be clustering a set of malicious domains into one cyber campaign.


\bibliographystyle{splncs03}
\bibliography{Bibliography}

\end{document}